\documentclass{article}
\usepackage{dcase2018,amsmath,graphicx,url,times,booktabs, tabularx}
\usepackage[final]{microtype}
\usepackage[T1]{fontenc}

\title{DCASE 2018 Challenge Surrey Cross-Task convolutional neural network baseline}

\name{Qiuqiang Kong$^1$, Turab Iqbal$^1$, Yong Xu$^2$, Wenwu Wang$^1$, Mark D. Plumbley$^1$}
\address{Centre for Vision, Speech and Signal Processing (CVSSP), University of Surrey \\
$^1$\{q.kong, t.iqbal, w.wang, m.plumbley\}@surrey.ac.uk \\
$^2$yong.xu.ustc@gmail.com}

\begin{document}

\ninept
\maketitle

\begin{sloppy}

\begin{abstract}
The Detection and Classification of Acoustic Scenes and Events (DCASE) consists of five audio classification and sound event detection tasks: 1) Acoustic scene classification, 2) General-purpose audio tagging of Freesound, 3) Bird audio detection, 4) Weakly-labeled semi-supervised sound event detection and 5) Multi-channel audio classification. In this paper, we create a cross-task baseline system for all five tasks based on a convlutional neural network (CNN): a ``CNN Baseline'' system. We implemented CNNs with 4 layers and 8 layers originating from AlexNet and VGG from computer vision. We investigated how the performance varies from task to task with the same configuration of neural networks. Experiments show that deeper CNN with 8 layers performs better than CNN with 4 layers on all tasks except Task 1. Using CNN with 8 layers, we achieve an accuracy of 0.680 on Task 1, an accuracy of 0.895 and a mean average precision (MAP) of 0.928 on Task 2, an accuracy of 0.751 and an area under the curve (AUC) of 0.854 on Task 3, a sound event detection F1 score of 20.8\% on Task 4, and an F1 score of 87.75\% on Task 5. We released the Python source code of the baseline systems under the MIT license for further research. 
\end{abstract}


\section{Introduction}

Detection and classification of acoustic scenes and events (DCASE) 2018 challenge\footnote{http://dcase.community/} is a well known IEEE challenge consists of several audio classification and sound event detection tasks. DCASE 2018 challenge consists of five tasks: In task 1, acoustic scene classification (ASC) \cite{dcase2018task1}, the task is to recognize the scenes where the sound is recorded, such as ``street'' or ``park''. ASC has applications in enhancing speech recognition systems and sound event detection \cite{barchiesi2014acoustic}. Task 1 includes a matching device ASC subtask and a mismatching device ASC subtask. In task 2, general-purpose audio tagging of Freesound, \cite{dcase2018task2} the task is to classify an audio clip to a pre-defined class, such as ``flute'' or ``applause''. Task 2 has applications in recognizing a wide range of sound events in real world and is useful for information retrieval. In task 3, bird audio detection, \cite{dcase2018task3}, the task is to detect the presence or the absence of birds in an audio clip. This could be used for automatic wildlife monitoring and audio library management. An important goal of Task 3 is to design a classification system which can generalize to new conditions. In task 4, weakly labeled semi-supervised sound event detection (SED) \cite{dcase2018task4}, the task is to detect the onset the offset times of sound events where only weak labeled audio and unlabeled audio is available for training. Task 4 can be used for monitoring public security and used for abnormal sound detection. In task 5, the multi-channel audio classification \cite{dcase2018task5}, the task is to use multi-channel recordings to identify the human activities at home. 

The first DCASE challenge was the DCASE 2013 challenge \cite{giannoulis2013detection}, with only an audio classification and a sound event detection tasks. The DCASE 2016 challenge \cite{mesaros2016tut} consisted of four tasks including: 1) ASC, 2) SED in synthetic audio, 3) SED in real audio and 4) domestic audio tagging. The DCASE 2017 challenge \cite{mesaros2017dcase} updated the domestic audio tagging task to a large-scale weakly labeled audio tagging task. The DCASE challenge series provide public datasets for investigating audio related tasks. One recent dataset for DCASE challenges is the AudioSet dataset \cite{gemmeke2017audio}. Task 4 of both DCASE 2017 and 2018 challenge were subsets of AudioSet. 

Convolutional neural networks (CNNs) have achieved state-of-the-art performance in image classification \cite{krizhevsky2012imagenet, simonyan2014very}. In this paper, we investigate how different CNNs including CNN with 4 layers originated from AlexNet \cite{krizhevsky2012imagenet} and CNN with 8 layers originated from VGG \cite{simonyan2014very} perform on Task 1 to 5 of DCASE 2018. We apply the same configurations of CNNs across all task 1 to 5 to fairly compare the relative performance across different tasks. Using the same CNN model, the performance on Task 1 to 5 varies, which indicates the difficulty of the tasks varies. The experiments show that Task 4 sound event detection is more difficult than Task 1 acoustic scene classification than Task 3 bird audio detection than Task 2 general-purpose audio tagging of Freesound and Task 5 domestic multi-channel audio tagging. 

We open source the Python code for all of Task 1 - 5 of DCASE 2018 challenge under MIT license. The source code contains the implementation of CNNs with 4 layers and 8 layers. In complementary to the source code published by the organizer \cite{dcase2018task1}, we investigated that CNNs with more layers perform better in all of Task 2 - 5 in DCASE 2018 challenge except Task 1. 

This paper is organized as follows, Section 2 introduces related works. Section 3 introduces CNNs. Section 4 shows experimental results. Section 5 concludes and forecasts our work. 

\section{Related works}
Manually-selected features such as mel frequency cepstrum coefficients (MFCC) \cite{li2001classification}, the constant Q transform (CQT) \cite{bisot2016supervised}, and I-vectors \cite{eghbal2016cp} have been used as audio features. Recently, mel spectrograms \cite{choi2016automatic} have been widely used as features when using neural networks as classifiers. Mixture Gaussian models (GMMs) \cite{stowell2015detection} and hidden Markov models (HMMs) \cite{mesaros2010acoustic} have been used to model audio scenes and sound events. Non-negative matrix factorization (NMFs) \cite{mesaros2015sound} are methods to learn a set of bases to represent the audio. Recently, deep neural networks have been introduced to audio classification and sound event detection. For example, fully-connected neural networks have been applied to DCASE 2016 challenges \cite{kong2016deep} and DCASE 2017 challenges \cite{li2017comparison}. CNNs have achieved the state-of-the-art performance in audio classification and sound event detection \cite{hershey2017cnn, choi2016automatic, aytar2016soundnet}. Convolutional recurrent neural networks (RNNs) \cite{cakir2017convolutional, lim2017rare} have been used to model the temporal information of sound events. Attention neural networks have been proposed to focus on sound events \cite{xu2017large} from weakly-labelled data \cite{kumar2016audio}. Generative adversarial networks (GANs) have been applied to improve the robustness of audio classification classifiers \cite{mun2017generative}. 

\begin{table}[t]
\centering
\caption{Configurations of CNN4 and CNN8}
\label{tab:parameter}
\begin{tabular}{c|c|c}
\hline
\begin{tabular}[c]{@{}l@{}}feature map\\ size\end{tabular} & CNN4          & \multicolumn{1}{c}{CNN8}                                                                                   \\ \hline
$ T \times 64 $                                            & \multicolumn{2}{c}{log mel spectrogram}                                                                                            \\ \hline
{$ T / 2 \times 32 $}                       & $ 5 \times 5, 64 $  & \begin{tabular}[c]{@{}l@{}}$ \begin{bmatrix} 3 \times 3, \text{BN} \\ 3 \times 3, \text{BN} \end{bmatrix}, 64 $\end{tabular}  \\ \cline{2-3} 
                                                           & \multicolumn{2}{c}{$ 2 \times 2 $, max pooling}                                                                                    \\ \hline
{$ T / 4 \times 16 $}                       & $ 5 \times 5, 128 $ & \begin{tabular}[c]{@{}l@{}}$ \begin{bmatrix} 3 \times 3, \text{BN} \\ 3 \times 3, \text{BN} \end{bmatrix}, 128 $\end{tabular} \\ \cline{2-3} 
                                                           & \multicolumn{2}{c}{$ 2 \times 2 $, max pooling}                                                                                    \\ \hline
{$ T / 8 \times 8 $}                        & $ 5 \times 5, 256 $ & \begin{tabular}[c]{@{}l@{}}$ \begin{bmatrix} 3 \times 3, \text{BN} \\ 3 \times 3, \text{BN} \end{bmatrix}, 256 $\end{tabular} \\ \cline{2-3} 
                                                           & \multicolumn{2}{c}{$ 2 \times 2 $, max pooling}                                                                                    \\ \hline
{$ T / 16 \times 4 $}                       & $ 5 \times 5, 512 $ & \begin{tabular}[c]{@{}l@{}}$ \begin{bmatrix} 3 \times 3, \text{BN} \\ 3 \times 3, \text{BN} \end{bmatrix}, 512 $\end{tabular} \\ \cline{2-3} 
                                                           & \multicolumn{2}{c}{$ 2 \times 2 $, max pooling}                                                                                    \\ \hline
$ 1 \times 1 $                                             & \multicolumn{2}{c}{Global max pooling}                                                                                             \\ \hline
                                                           & \multicolumn{2}{c}{Classes num. fc, sigmoid or softmax}                                                                              \\ \hline
Parameters & 4,309,450 & 4,691,274 \\ \hline
\end{tabular}
\end{table}

\section{Convolutional neural networks}

CNNs, such as AlexNet \cite{krizhevsky2012imagenet} and VGG \cite{simonyan2014very}, have achieved state-of-the-art performance in image classification \cite{krizhevsky2012imagenet, simonyan2014very}. A CNN consists of several convolutional layers followed by fully-connected layers. Each convolutional layer consists of filters to convolve with the output from the previous convolutional layer. The filters can capture local patterns in feature maps, such as edges in lower layers and complex profiles in higher layers \cite{simonyan2014very}. In this work, we adopt AlexNet with 4 layers and VGG with 8 layers as models, which we call CNN4 and CNN8. CNN4 consists of 4 convolutional layers and the filter size of each convolutional layer is $ 5 \times 5 $ \cite{krizhevsky2012imagenet}. CNN8 consists of 8 layers and the filter size of each convolutional layer is $ 3 \times 3 $ \cite{simonyan2014very}. We apply batch normalization (BN) after each convolutional layer to stabilize training \cite{ioffe2015batch} followed by a rectifier (ReLU) nonlinearity. We then apply a global max pooling (GMP) operation on the feature maps of the last convolutional layer \cite{choi2016automatic} to summarize the feature maps to a vector. GMP can max out the time and frequency information of sound events in a spectrogram, so it is invariant to time or frequency shift. Finally, a fully-connected layer is applied on the summarized vector followed by a sigmoid or softmax nonlinearity to output the probabilities of the audio classes. The configurations of CNN4 and CNN8 are summarized in Table \ref{tab:parameter}.

\begin{table}[t]
  \caption{Task 1 acoustic scene classification class-wise accuracy of subtask A and B of development dataset. }
  \vspace{6pt}
  \label{tab:task1}
  \centering
  \begin{tabular}{l  p{.6cm}p{.6cm}p{.6cm}p{.6cm}p{.6cm}p{.6cm}}
    \toprule
    & \multicolumn{3}{c}{\textbf{\textsc{Subtask A}}} & \multicolumn{3}{c}{\textbf{\textsc{Subtask B}}} \\
	\cmidrule(lr){2-4} \cmidrule(lr){5-7} 
    Scene label & CNN \cite{dcase2018task1} & CNN4 & CNN8 & CNN \cite{dcase2018task1} & CNN4 & CNN8 \\
    \midrule
    Airport            & 0.729 & 0.743 & 0.709 & 0.725 & 0.612 & 0.667 \\
    Bus                & 0.629 & 0.607 & 0.649 & 0.783 & 0.695 & 0.723 \\
    Metro              & 0.512 & 0.690 & 0.686 & 0.206 & 0.500 & 0.417 \\
    Metro station      & 0.554 & 0.687 & 0.741 & 0.328 & 0.472 & 0.584 \\
    Park               & 0.791 & 0.855 & 0.839 & 0.592 & 0.834 & 0.861 \\
    public square      & 0.404 & 0.486 & 0.472 & 0.247 & 0.361 & 0.389 \\
    Shopping mall      & 0.496 & 0.642 & 0.631 & 0.611 & 0.778 & 0.778 \\
    Street, pedestrian & 0.500 & 0.583 & 0.567 & 0.208 & 0.333 & 0.361 \\
    Street, traffic    & 0.805 & 0.874 & 0.886 & 0.664 & 0.750 & 0.778 \\
    Tram               & 0.551 & 0.590 & 0.621 & 0.197 & 0.417 & 0.389 \\
    \midrule
    \textbf{Average}   & 0.597 & 0.676 & \textbf{0.680} & 0.456 & \textbf{0.575} & 0.572 \\
    \midrule
    \textbf{Public LB} & - & 0.693 & \textbf{0.707} & - & \textbf{0.578} & 0.568 \\
    \midrule
    \textbf{Private LB} & - & 0.628 & \textbf{0.630} & - & 0.615 & \textbf{0.672} \\
    \midrule
    \textbf{Evaluation} & - & 0.697 & \textbf{0.704} & - & 0.588 & \textbf{0.596} \\
	\bottomrule
\end{tabular}
\end{table}

\section{Experiments}

We open source the Python code of the CNN baseline systems of DCASE 2018 challenge Task 1 - 5 source here\footnote{https://github.com/qiuqiangkong/dcase2018\textunderscore task1}\footnote{https://github.com/qiuqiangkong/dcase2018\textunderscore task2}\footnote{https://github.com/qiuqiangkong/dcase2018\textunderscore task3}\footnote{https://github.com/qiuqiangkong/dcase2018\textunderscore task4}\footnote{https://github.com/qiuqiangkong/dcase2018\textunderscore task5}. We convert all stereo audio to mono for Task 1 - 5 for building the baseline system. We extract the spectrograms and apply log mel filter banks on the spectrograms followed by logarithm operation. We choose the number of the mel filter banks as 64 because it is a power of two which can be divided by two in max pooling layers. The mel filter bank has a cut off frequency of 50 Hz. The log mel spectrograms are standarized by subtracting the mean and dividing the standard deviation along mel frequency bins. The same configuration of CNN4 and CNN8 are applied on Task 1 - 5. We use Adam optimizer \cite{kingma2014adam} with a learning rate of 0.001 and the learning rate is reduced by multiplying 0.9 after every 200 iterations training. A batch size of 128 is used for Task 1, 2, 3 and 5 and a batch size of 32 is used for Task 4 to sufficiently use the GPU with 12 GB memory in training. We trained the model for 5000 iterations for all of the five tasks. The training takes 60 ms and 200 ms per iteration on a Titan X GPU for CNN4 and CNN8, respectively. The results of Task 1 - 5 are shown in the following subsections. 

\subsection{Task 1: Acoustic scene classification}

Task 1 acoustic scene classification \cite{dcase2018task1} is a task to classify an audio recording to a predefined class that characterize the environment in which it was recorded. The 10 predefined classes are listed in Table \ref{tab:task1}. There are 10080 10-second audio clips in the development dataset, including 8640, 720 and 720 audio clips recorded with device A, B and C. Task 1 has three subtasks. Subtask A is matching device classification. Subtask B is mismatching device classification. Subtask C is matching device classification with external data and has the same evaluation data as subtask A. 

Table \ref{tab:task1} shows the accuracy of subtask A and subtask B. In \cite{dcase2018task1} a two layer CNN with a dense connected layer is used as a baseline model. In development dataset of subtask A, CNN4 and CNN8 achieve similar accuracy of 0.676 and 0.680 respectively, outperforming the two layers CNN of 0.597 \cite{dcase2018task1}. In subtask B, CNN4 and CNN8 achieve similar accuracy of 0.575 and 0.572, respectively, outperforming the two layers CNN of 0.456 \cite{dcase2018task1}. The subtask B mismatching device classification is around 10\% which is worse than the subtask A matching device classification in absolute value. Table \ref{tab:task1} also shows the public leaderboard (LB), private LB and final evaluation result. We did not explore the subtask C with external data. 

\begin{table}
  \caption{Task 2 audio tagging accuracy and MAP@3. }
  \vspace{6pt}
  \label{tab:task2}
  \centering
  \begin{tabular}{l p{0.8cm}p{0.8cm}p{0.8cm}p{0.8cm}}
    \toprule
    & \multicolumn{2}{c}{\textbf{Accuracy}} & \multicolumn{2}{c}{\textbf{MAP@3}} \\
	\cmidrule(lr){2-3} \cmidrule(lr){4-5}
     & CNN4 & CNN8 & CNN4 & CNN8 \\
    \midrule
    Fold 1 & 0.858 & 0.897 & 0.900 & 0.930 \\
    Fold 2 & 0.824 & 0.875 & 0.870 & 0.912 \\
    Fold 3 & 0.862 & 0.903 & 0.901 & 0.934 \\
    Fold 4 & 0.861 & 0.904 & 0.904 & 0.935 \\
    \midrule
    \textbf{Average} & 0.851 & \textbf{0.895} & 0.894 & \textbf{0.928} \\
    \midrule
    \textbf{Public LB} & - & - & 0.885 & \textbf{0.920} \\
    \midrule 
    \textbf{Private LB} & - & - & 0.862 & \textbf{0.903} \\
	\bottomrule
\end{tabular}
\end{table}

\begin{figure}[t]
  \centering
  \centerline{\includegraphics[width=\columnwidth]{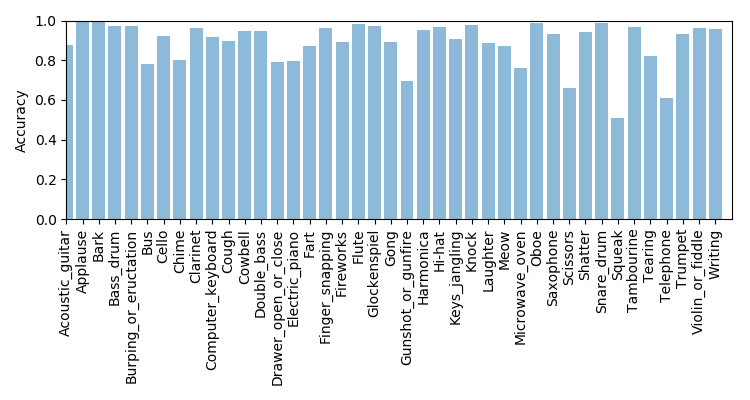}}
  \caption{Task 2 audio tagging class-wise accuracy. }
  \label{fig:task2}
\end{figure}

\subsection{Task 2: General-purpose audio tagging of Freesound content with AudioSet labels}

Task 2 audio tagging \cite{dcase2018task2} is a task to classify an audio clip to one of 41 predefined classes such as ``oboe'' and ``applause''. The duration of the audio samples range from 300 ms to 30 s due to the diversity of the sound categories. The development dataset contains 9473 audio clips. We pad or split the log mel spectrograms of audio clips to 2 s log mel spectrograms as the input to a CNN. We split the development dataset to four validation folds and only use 3710 manually verified audio clips for validation. Table \ref{tab:task2} shows the accuracy and the mean average precision (MAP) \cite{dcase2018task2} on the four folds and their average statistics. In development dataset, CNN8 achieves an average accuracy of 0.895 and a MAP@3 of 0.928, outperforming CNN4 network of 0.851 and 0.894, respectively. Figure \ref{fig:task2} shows the averaged 4 folds class-wise accuracy of Task 2. Sound classes such as ``applause'' and ``bark'' have 100\% classification accuracy but some classes such as ``squeak'' and ``telephone'' have accuracy of only 50\% - 60\%. Table \ref{tab:task2} shows the MAP@3 of the private leaderboard is approximately 2\% worse than the development and the public leaderboard. 
\begin{table}
  \caption{Task 3 bird audio detection accuracy and AUC.  }
  \vspace{6pt}
  \label{tab:task3}
  \centering
  \begin{tabular}{l p{0.8cm}p{0.8cm}p{0.8cm}p{0.8cm}}
    \toprule
    & \multicolumn{2}{c}{\textbf{Accuracy}} & \multicolumn{2}{c}{\textbf{AUC}} \\
	\cmidrule(lr){2-3} \cmidrule(lr){4-5}
    validation dataset & CNN4 & CNN8 & CNN4 & CNN8 \\
    \midrule
    freefield1010 & 0.551 & 0.630 & 0.645 & 0.799 \\
    warblrb10k & 0.692 & 0.867 & 0.799 & 0.882 \\
    BirdVox-DCASE-20K & 0.678 & 0.801 & 0.808 & 0.882 \\
    \midrule
    \textbf{Average} & 0.640 & \textbf{0.766} & 0.751 & \textbf{0.854} \\
    \midrule
    \textbf{Leaderboard} & - & - & \textbf{0.850} & 0.847 \\
    \midrule
    \textbf{Evaluation} & - & - & 0.748 & \textbf{0.809} \\
	\bottomrule
\end{tabular}
\end{table}

\begin{table}[b!]
  \caption{Task 4 audio tagging AUC and sound event detection F1 score. }
  \vspace{6pt}
  \label{tab:task4}
  \centering
  \begin{tabular}{l p{0.6cm}p{0.6cm}p{.6cm}p{.6cm}p{0.6cm}p{0.6cm}}
    \toprule
    & \multicolumn{2}{c}{\textbf{AT (AUC)}} & \multicolumn{2}{c}{\textbf{SED1 (F1)}} & \multicolumn{2}{c}{\textbf{SED2 (F1)}} \\
	\cmidrule(lr){2-3} \cmidrule(lr){4-5} \cmidrule(lr){6-7}
    Class                       & CNN4 & CNN8 & CNN4 & CNN8 & CNN4 & CNN8 \\
    \midrule
    Speech                      & 0.889 & 0.936 & 0.0\% & 0.0\% & 16.9\% & \textbf{22.5\%} \\ 
    Dog                         & 1.000 & 1.000 & 2.6\% & 2.5\% & 8.3\% & \textbf{14.3\%} \\
    Cat                         & 0.980 & 0.991 & 3.4\% & 3.5\% & \textbf{10.3\%} & 7.2\% \\
    Alarm/bell           & 0.964 & 0.975 & 4.2\% & 4.0\% & 12.5\% & \textbf{20.7\%} \\
    Dishes                      & 0.835 & 0.898 & 0.0\% & 0.0\% & 0.0\% & \textbf{3.6\%} \\
    Frying                      & 0.945 & 0.939 & 45.5\% & \textbf{54.5\%} & 2.1\% & 0.0\% \\
    Blender                     & 0.839 & 0.883 & 18.9\% & \textbf{27.1\%} & 8.3\% & 7.3\% \\
    Running water               & 0.930 & 0.943 & 11.8\% & \textbf{11.9\%} & 7.9\% & 3.1\% \\
    Vacuum cleaner              & 0.972 & 0.956 & 57.6\% & \textbf{61.3\%} & 9.4\% & 2.6\% \\
    Electronic shaver & 0.944 & 0.957 & \textbf{45.0\%} & 43.5\% & 18.9\% & 16.3\% \\
    \midrule
    \textbf{Average}                     & 0.930 & 0.948 & 18.9\% & \textbf{20.8\%} & 9.5\% & 9.8\% \\
    \midrule
    \textbf{Evaluation} & - & - & 16.7\% & \textbf{18.6\%} & - & - \\
	\bottomrule
\end{tabular}
\end{table}

\begin{table*}[t]
  \caption{Task 5 multi-channel audio tagging F1 score. }
  \vspace{6pt}
  \label{tab:task5}
  \centering
  \begin{tabular}{l p{1cm}p{0.8cm}p{0.8cm}p{.8cm}p{.8cm}p{1cm}p{.8cm}p{.8cm}p{.8cm}p{.8cm}p{1cm}}
    \toprule
    & & \multicolumn{5}{c}{\textbf{CNN4 (F1 score)}} & \multicolumn{5}{c}{\textbf{CNN8 (F1 score)}}  \\
	\cmidrule(lr){3-7} \cmidrule(lr){8-12}
    Scene label      & Baseline & Fold 1 & Fold 2 & Fold 3 & Fold 4 & Average & Fold 1 & Fold 2 & Fold 3 & Fold 4 & Average \\
    \midrule
    Absence          & 85.4\% & 86.4\% & 90.5\% & 78.5\% & 89.9\% & 86.3\% & 90.5\% & 92.2\% & 80.5\% & 89.9\% & 88.3\% \\
    Cooking          & 95.1\% & 96.2\% & 94.7\% & 93.0\% & 96.6\% & 95.1\% & 98.0\% & 96.3\% & 93.8\% & 96.3\% & 96.1\% \\
    Dishwashing      & 76.7\% & 77.8\% & 68.6\% & 75.8\% & 80.2\% & 75.6\% & 83.3\% & 71.2\% & 76.0\% & 85.8\% & 79.1\% \\
    Eating           & 83.6\% & 79.7\% & 75.7\% & 85.4\% & 91.2\% & 82.3\% & 85.2\% & 85.1\% & 88.5\% & 94.5\% & 88.3\% \\
    Other            & 44.8\% & 43.3\% & 55.2\% & 56.9\% & 60.2\% & 53.9\% & 54.3\% & 54.5\% & 51.4\% & 62.2\% & 55.6\% \\
    Social activity  & 93.9\% & 95.5\% & 88.1\% & 90.2\% & 98.5\% & 93.1\% & 98.4\% & 90.1\% & 93.7\% & 99.3\% & 95.4\% \\
    Vacuum cleaner   & 99.3\% & 100.0\% & 100.0\% & 100.0\% & 100.0\% & 100.0\% & 100.0\% & 99.4\% & 100.0\% & 100.0\% & 99.9\% \\
    Watching TV      & 99.6\% & 99.6\% & 99.7\% & 97.5\% & 100.0\% & 99.2\% & 99.8\% & 99.9\% & 99.0\% & 99.9\% & 99.7\% \\
    Working          & 82.0\% & 85.3\% & 86.3\% & 79.4\% & 90.5\% & 85.4\% & 88.7\% & 89.3\% & 81.4\% & 90.2\% & 87.4\% \\
    \midrule
    \textbf{Average}          & 84.5\% & 84.9\% & 84.3\% & 84.1\% & 89.7\% & \textbf{85.7\%} & 88.7\% & 86.5\% & 84.9\% & 90.9\% & \textbf{87.8\%} \\
    \midrule
    \textbf{Eval. Unknown mic.} & 83.1\% & - & - & - & - & 82.4\% & - & - & - & - & \textbf{83.2\%} \\
    \midrule
    \textbf{Eval. dev. mic.} & 85.0\% & - & - & - & - & 86.2\% & - & - & - & - & \textbf{87.6\%} \\
	\bottomrule
\end{tabular}
\end{table*}

\subsection{Task 3: Bird audio detection}

Task 3 bird audio detection \cite{dcase2018task3} is a task to predict the presence or the absence of birds in a 10-second audio clip. One challenge of this task is to design a system that is able to generalize to new conditions. That is, a system trained on one dataset should generalize well to another dataset. The development dataset consists of freefield1010 with 7690 audio clips, warblrb10k with 8000 audio clips and BirdVox-DCASE-20K with 20000 audio clips. We train on two datasets and evaluate on the other hold out dataset. Table \ref{tab:task3} shows the accuracy and the area under the curve (AUC) \cite{dcase2018task3} of CNN4 and CNN8. In development dataset, CNN8 achieves an accuracy and an AUC of 0.766 and 0.854, outperforming CNN4 of 0.640 and 0.751, respectively. The result in Table \ref{tab:task3} shows the classification of freefield1010 dataset is more difficult than warblrb10k and BirdVox-DCASE-20K dataset. Furthermore, an AUC of 0.809 is achieved in evaluation dataset using CNN8.

\subsection{Task 4: Large-scale weakly labeled semi-supervised sound event detection in domestic environments}

Task 4 is a weakly labeled semi-supervised sound event detection task \cite{dcase2018task4} to predict both the onset and offset of sound events. There are 10 audio classes in Task 4, for example ``speech'' and ``dog''. An audio clip can be assigned to one or more labels. The development dataset consists of 1578 weakly labeled audio clips, 14412 unlabeled in domain audio clips and 39999 unlabeled out domain audio clips. Each audio clip has a duration of 10 seconds. We only use the 1578 weakly labeled audio clips for training our systems. Different from Task 1, 2, 3 and 5, to remain the time resolution of feature maps in time axis, max pooling operations are only applied along the frequency axis but not the time axis. In training, we average out the time axis and apply a fully connected layer to predict the clip-wise labels. In inference, we do not apply the average of time axis to remain frame-wise labels. Table \ref{tab:task4} shows CNN8 achieves an AUC of 0.948 in audio tagging, outperforming CNN4 of 0.930. In sound event detection, system SED1 uses the audio tagging result as the sound event detection result. The onset and offset times are filled with 0 s and 10 s. System SED2 applies thresholds to the frame-wise predictions to detect sound events. The high threshold and the low threshold are set as 0.8 and 0.2, respectively. Sound events such as ``Frying'', ``Blender'' have higher F1 score with SED1. Sound event such as ``Speech'', ``Dog'', ``Cat'' have higher F1 score with SED2. In development dataset, SED1 and SED2 achieve average F1 scores of 20.8\% and 9.8\%, respectively. In evaluation, a F1 score of 18.6\% is achieved using CNN8 and system SED1.

\subsection{Task 5: Monitoring of domestic activities based on multi-channel acoustics}

Task 5 multi-channel audio tagging \cite{dcase2018task5} is a task to classify the domestic activities with multi-channel acoustic recordings. The target of Task 5 is to research how the multi-channel information will help the audio tagging task. The development dataset of Task 5 consists of 72984 10-second audio clips. The audio classes including ``Cooking'' and ``Eating'', for example. The multi-channel audio clips are converted to single channel audio clips to build the baseline system. Table \ref{tab:task5} shows in development dataset the CNN8 achieves a F1 score of 87.75\%, outperforming CNN4 network of 85.73\%. In Evaluation data with unknown microphone a F1 score of 83.2\% is achieved using CNN8 model. With unkown development microphone, a F1 score of 87.6\% is achieved.

\section{Conclusion}

In this paper, we investigated the performance of convolutional neural networks (CNNs) with 4 layers and 8 layers on Task 1 to 5 of DCASE 2018. We show the difficulties of the tasks varies. Task 4 sound event detection is more difficult than Task 1 acoustic scene classification than Task 3 bird audio detection than Task 2 general-purpose audio tagging of Freesound and Task 5 domestic multi-channel audio tagging. We show CNN with 8 layers performs better than CNN with 4 layers in Task 2 to 5. In Task 1, CNN with 8 layers and 4 layers perform similar. With CNN8, we achieves an accuracy of 0.680 on Task 1, a mean average precision (MAP) of 0.928 on Task 2, an area under the curve (AUC) of 0.854 on Task 3, a sound event detection F1 score of 20.8\% on Task 4 and a F1 score of 87.75\% on Task 5. In future, we will explore more CNN structures on Task 1 to 5 of DCASE 2018 challenge. We released the Python source code of the baseline systems under the MIT license for further research. 

\section{Acknowledgement}
This research was supported by EPSRC grant EP/N014111/1 ``Making Sense of Sounds'' and a Research Scholarship from the China Scholarship Council (CSC) No. 201406150082.

\bibliographystyle{IEEEtran}
\bibliography{refs}

\begin{thebibliography}{10}
\providecommand{\url}[1]{#1}
\csname url@samestyle\endcsname
\providecommand{\newblock}{\relax}
\providecommand{\bibinfo}[2]{#2}
\providecommand{\BIBentrySTDinterwordspacing}{\spaceskip=0pt\relax}
\providecommand{\BIBentryALTinterwordstretchfactor}{4}
\providecommand{\BIBentryALTinterwordspacing}{\spaceskip=\fontdimen2\font plus
\BIBentryALTinterwordstretchfactor\fontdimen3\font minus
  \fontdimen4\font\relax}
\providecommand{\BIBforeignlanguage}[2]{{%
\expandafter\ifx\csname l@#1\endcsname\relax
\typeout{** WARNING: IEEEtran.bst: No hyphenation pattern has been}%
\typeout{** loaded for the language `#1'. Using the pattern for}%
\typeout{** the default language instead.}%
\else
\language=\csname l@#1\endcsname
\fi
#2}}
\providecommand{\BIBdecl}{\relax}
\BIBdecl

\bibitem{dcase2018task1}
A.~Mesaros, T.~Heittola, and T.~Virtanen, ``A multi-device dataset for urban
  acoustic scene classification,'' \emph{arXiv preprint arXiv:1807.09840},
  2018.

\bibitem{barchiesi2014acoustic}
D.~Barchiesi, D.~Giannoulis, D.~Stowell, and M.~D. Plumbley, ``Acoustic scene
  classification,'' \emph{arXiv preprint arXiv:1411.3715}, 2014.

\bibitem{dcase2018task2}
E.~Fonseca, M.~Plakal, F.~Font, D.~P.~W. Ellis, X.~Favory, J.~Pons, and
  X.~Serra, ``General-purpose tagging of freesound audio with audioset labels:
  Task description, dataset, and baseline,'' \emph{arXiv preprint
  arXiv:1807.09902}, 2018.

\bibitem{dcase2018task3}
D.~Stowell, Y.~Stylianou, M.~Wood, H.~Pamu{\l}a, and H.~Glotin, ``Automatic
  acoustic detection of birds through deep learning: the first bird audio
  detection challenge,'' \emph{arXiv preprint arXiv:1807.05812}, 2018.

\bibitem{dcase2018task4}
R.~Serizel, T.~Nicolas, H.~Eghbal-Zadeh, and A.~P. Shah, ``Large-scale weakly
  labeled semi-supervised sound event detection in domestic environments,''
  \emph{https://hal.inria.fr/hal-01850270}, 2018.

\bibitem{dcase2018task5}
G.~Dekkers, S.~Lauwereins, B.~Thoen, M.~W. Adhana, H.~Brouckxon, T.~van
  Waterschoot, B.~Vanrumste, M.~Verhelst, and P.~Karsmakers, ``The {SINS}
  database for detection of daily activities in a home environment using an
  acoustic sensor network,'' in \emph{Proceedings of the Detection and
  Classification of Acoustic Scenes and Events 2017 Workshop (DCASE2017)},
  Munich, Germany, November 2017, pp. 32--36.

\bibitem{giannoulis2013detection}
D.~Giannoulis, E.~Benetos, D.~Stowell, M.~Rossignol, M.~Lagrange, and M.~D.
  Plumbley, ``Detection and classification of acoustic scenes and events: {An
  IEEE AASP} challenge,'' in \emph{IEEE Workshop on Applications of Signal
  Processing to Audio and Acoustics (WASPAA)}.\hskip 1em plus 0.5em minus
  0.4em\relax IEEE, 2013, pp. 1--4.

\bibitem{mesaros2016tut}
A.~Mesaros, T.~Heittola, and T.~Virtanen, ``{TUT} database for acoustic scene
  classification and sound event detection,'' in \emph{Signal Processing
  Conference (EUSIPCO)}.\hskip 1em plus 0.5em minus 0.4em\relax IEEE, 2016, pp.
  1128--1132.

\bibitem{mesaros2017dcase}
A.~Mesaros, T.~Heittola, A.~Diment, B.~Elizalde, A.~Shah, E.~Vincent, B.~Raj,
  and T.~Virtanen, ``{DCASE 2017 challenge setup: Tasks, datasets and baseline
  system},'' in \emph{DCASE 2017-Workshop on Detection and Classification of
  Acoustic Scenes and Events}, 2017.

\bibitem{gemmeke2017audio}
J.~F. Gemmeke, D.~P.~W. Ellis, D.~Freedman, A.~Jansen, W.~Lawrence, R.~C.
  Moore, M.~Plakal, and M.~Ritter, ``Audio set: An ontology and human-labeled
  dataset for audio events,'' in \emph{IEEE International Conference on
  Acoustics, Speech and Signal Processing (ICASSP)}.\hskip 1em plus 0.5em minus
  0.4em\relax IEEE, 2017, pp. 776--780.

\bibitem{krizhevsky2012imagenet}
A.~Krizhevsky, I.~Sutskever, and G.~E. Hinton, ``Imagenet classification with
  deep convolutional neural networks,'' in \emph{Advances in Neural Information
  Processing Systems (NIPS)}, 2012, pp. 1097--1105.

\bibitem{simonyan2014very}
K.~Simonyan and A.~Zisserman, ``Very deep convolutional networks for
  large-scale image recognition,'' \emph{arXiv preprint arXiv:1409.1556}, 2014.

\bibitem{li2001classification}
D.~Li, I.~K. Sethi, N.~Dimitrova, and T.~McGee, ``Classification of general
  audio data for content-based retrieval,'' \emph{Pattern Recognition Letters},
  vol.~22, no.~5, pp. 533--544, 2001.

\bibitem{bisot2016supervised}
V.~Bisot, R.~Serizel, S.~Essid, and G.~Richard, ``Supervised nonnegative matrix
  factorization for acoustic scene classification,'' \emph{IEEE AASP Challenge
  on Detection and Classification of Acoustic Scenes and Events (DCASE)}, 2016.

\bibitem{eghbal2016cp}
H.~Eghbal-Zadeh, B.~Lehner, M.~Dorfer, and G.~Widmer, ``{CP-JKU submissions for
  DCASE-2016: A hybrid approach using binaural i-vectors and deep convolutional
  neural networks},'' \emph{IEEE AASP Challenge on Detection and Classification
  of Acoustic Scenes and Events (DCASE)}, 2016.

\bibitem{choi2016automatic}
K.~Choi, G.~Fazekas, and M.~Sandler, ``Automatic tagging using deep
  convolutional neural networks,'' \emph{arXiv preprint arXiv:1606.00298},
  2016.

\bibitem{stowell2015detection}
D.~Stowell, D.~Giannoulis, E.~Benetos, M.~Lagrange, and M.~D. Plumbley,
  ``Detection and classification of acoustic scenes and events,'' \emph{IEEE
  Transactions on Multimedia}, vol.~17, no.~10, pp. 1733--1746, 2015.

\bibitem{mesaros2010acoustic}
A.~Mesaros, T.~Heittola, A.~Eronen, and T.~Virtanen, ``Acoustic event detection
  in real life recordings,'' in \emph{Signal Processing Conference}.\hskip 1em
  plus 0.5em minus 0.4em\relax IEEE, 2010, pp. 1267--1271.

\bibitem{mesaros2015sound}
A.~Mesaros, T.~Heittola, O.~Dikmen, and T.~Virtanen, ``Sound event detection in
  real life recordings using coupled matrix factorization of spectral
  representations and class activity annotations,'' in \emph{IEEE International
  Conference on Acoustics, Speech and Signal Processing (ICASSP)}.\hskip 1em
  plus 0.5em minus 0.4em\relax IEEE, 2015, pp. 151--155.

\bibitem{kong2016deep}
Q.~Kong, I.~Sobieraj, W.~Wang, and M.~D. Plumbley, ``Deep neural network
  baseline for dcase challenge 2016,'' \emph{Proceedings of DCASE 2016}, 2016.

\bibitem{li2017comparison}
J.~Li, W.~Dai, F.~Metze, S.~Qu, and S.~Das, ``A comparison of deep learning
  methods for environmental sound,'' \emph{arXiv preprint arXiv:1703.06902},
  2017.

\bibitem{hershey2017cnn}
S.~Hershey, S.~Chaudhuri, D.~P. Ellis, J.~F. Gemmeke, A.~Jansen, R.~C. Moore,
  M.~Plakal, D.~Platt, R.~A. Saurous, B.~Seybold \emph{et~al.}, ``{CNN
  architectures for large-scale audio classification},'' in \emph{IEEE
  International Conference on Acoustics, Speech and Signal Processing
  (ICASSP)}.\hskip 1em plus 0.5em minus 0.4em\relax IEEE, 2017, pp. 131--135.

\bibitem{aytar2016soundnet}
Y.~Aytar, C.~Vondrick, and A.~Torralba, ``Soundnet: Learning sound
  representations from unlabeled video,'' in \emph{Advances in Neural
  Information Processing Systems (NIPS)}, 2016, pp. 892--900.

\bibitem{cakir2017convolutional}
E.~Cakir, G.~Parascandolo, T.~Heittola, H.~Huttunen, T.~Virtanen, E.~Cakir,
  G.~Parascandolo, T.~Heittola, H.~Huttunen, and T.~Virtanen, ``Convolutional
  recurrent neural networks for polyphonic sound event detection,''
  \emph{IEEE/ACM Transactions on Audio, Speech and Language Processing
  (TASLP)}, vol.~25, no.~6, pp. 1291--1303, 2017.

\bibitem{lim2017rare}
H.~Lim, J.~Park, K.~Lee, and Y.~Han, ``{Rare sound event detection using 1D
  convolutional recurrent neural networks},'' DCASE2017 Challenge, Tech. Rep.,
  2017.

\bibitem{xu2017large}
Y.~Xu, Q.~Kong, W.~Wang, and M.~D. Plumbley, ``Large-scale weakly supervised
  audio classification using gated convolutional neural network,'' \emph{arXiv
  preprint arXiv:1710.00343}, 2017.

\bibitem{kumar2016audio}
A.~Kumar and B.~Raj, ``Audio event detection using weakly labeled data,'' in
  \emph{Proceedings of the 2016 ACM on Multimedia Conference}.\hskip 1em plus
  0.5em minus 0.4em\relax ACM, 2016, pp. 1038--1047.

\bibitem{mun2017generative}
S.~Mun, S.~Park, D.~K. Han, and H.~Ko, ``Generative adversarial network based
  acoustic scene training set augmentation and selection using svm
  hyper-plane,'' \emph{Proc. DCASE}, pp. 93--97, 2017.

\bibitem{ioffe2015batch}
S.~Ioffe and C.~Szegedy, ``Batch normalization: Accelerating deep network
  training by reducing internal covariate shift,'' \emph{arXiv preprint
  arXiv:1502.03167}, 2015.

\bibitem{kingma2014adam}
D.~P. Kingma and J.~Ba, ``Adam: A method for stochastic optimization,''
  \emph{arXiv preprint arXiv:1412.6980}, 2014.

\end{thebibliography}
%
%
%
%
%
%
%
%
%

\end{sloppy}
\end{document}